\documentclass[10pt,conference,IEEE]{IEEEtran}

\usepackage{algorithm}
\usepackage{algorithmic}
\usepackage{amsfonts}
\usepackage{amsmath}
\usepackage{bm}
\usepackage[usenames,dvipsnames]{color}

\usepackage[dvips]{graphicx}
\usepackage[update,prepend]{epstopdf}
\usepackage{graphicx}

\usepackage{caption}
\usepackage{subcaption}

\usepackage{url}
\usepackage{makeidx}
\usepackage{subfig}
\usepackage{siunitx}
\usepackage{sidecap}
\usepackage{latexsym}
\usepackage{amssymb}
\usepackage{units}
\usepackage{wasysym}

\ifCLASSOPTIONcompsoc
          \usepackage[nocompress]{cite}

\else

          \usepackage{cite}
\fi

\usepackage[acronym,nomain,nonumberlist,style=list]{glossaries}
\newacronym{DCR}{DCR}{Direct Conversion Receiver}
\newacronym{IC}{IC}{Integrated Circuit}
\newacronym{CMOS}{CMOS}{Complementary Metal--Oxide Semiconductor}
\newacronym{ADC}{ADC}{Analog-to-Digital Conversion}
\newacronym{AGC}{AGC}{Automatic Gain Control}
\newacronym{LO}{LO}{Local Oscillator}
\newacronym{SDR}{SDR}{Software Defined Radio}
\newacronym{COTS}{COTS}{Commercial Off-The Shelf}
\newacronym{FPGA}{FPGA}{Field Programmable Gate Array}
\newacronym{RISC}{RISC}{Reduced Instruction Set Computing}

\title{Compressed Sensing Based Direct Conversion Receiver
With Interference Reducing Sampling} 

\author{Jacek Pierzchlewski, Thomas Arildsen, Torben Larsen\\
Aalborg University,
Faculty of Engineering and Science, \\
Department of Electronic Systems,
Niels Jernes Vej 12, 9000 Aalborg, Denmark\\
jap@es.aau.dk, tha@es.aau.dk, tl@es.aau.dk\\
}

\begin{document}
\onecolumn
\Huge
This work has been submitted to the IEEE for possible publication.
Copyright may be transferred without notice, after which this version may no longer be accessible

\twocolumn
\maketitle
\thispagestyle{empty}

\begin{abstract}
This paper describes a direct conversion receiver applying compressed sensing
with the objective to relax the analog filtering requirements seen in the traditional
architecture.  The analog filter is cumbersome in an \gls{IC} design and relaxing
its requirements is an advantage in terms of die area, performance and robustness
of the receiver.  The objective is met by a selection of sampling pattern matched
to the prior knowledge of the frequency placement of the desired and interfering
signals.  A simple numerical example demonstrates the principle. The work is
part of an ongoing research effort and the different project phases are explained.
\end{abstract}

\section{Introduction}
\normalsize
The \gls{DCR} architecture is the most widely used receiver architecture
in mobile communication devices \cite{Abi01}.  This is caused by a small
analog part as well as reduced analog high quality filtering that causes
financially expensive and power and area inefficient off-chip interconnects.
Thus the \gls{DCR} is suitable for \gls{IC} implementation.
Nevertheless, there is a trend in the development of mobile receivers
to digitize even more of the receiver \cite{Abi01}.
\Glspl{DCR} are normally implemented in \gls{CMOS} which is excellent for digital
designs but problematic for analog designs due to high substrate losses and
tolerances making high-quality filtering very difficult if not impossible.
Therefore, it is generally preferred to have architectures with reduced
filtering demands and higher demands on the digital signal processing.

\begin{figure}[!]
            \centering
            \includegraphics{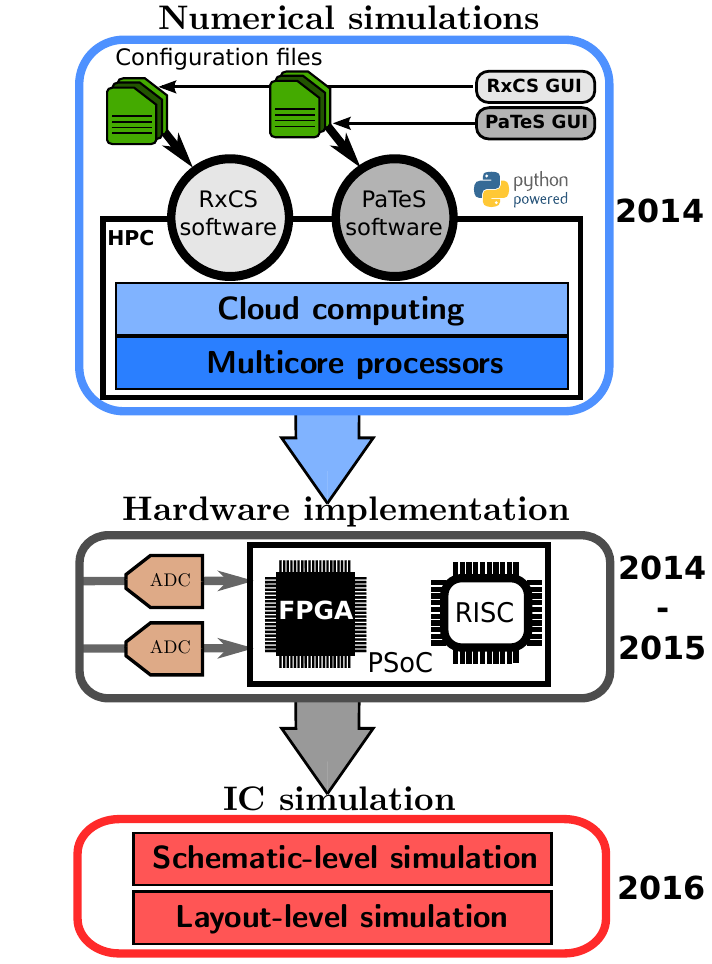}
            \caption{Phases of the IRfDUCS project.}
            \label{fig:system}
\end{figure}

In a \gls{DCR} the baseband signal(s) must be analog low-pass filtered
due to possible adjacent channel interference signals and to avoid
aliasing in the following \gls{ADC}.  The filter may also serve as partial
channel selection filter jointly with a digital filter.  If the analog filter does not significantly
reduce the adjacent interfering signal it adds to the requirement
of dynamic range and/or the sampling frequency of the \gls{ADC}
 \cite{Le05}.  One possible solution to reducing the analog filtering
 requirements is to use the principles of compressed sensing
\cite{Cand01} which allows signal acquisition with a sampling
frequency lower that the Nyquist frequency of a signal and also
allows usage of prior knowledge in the signal reconstruction.  In
this receiver case we may apply prior knowledge of the frequency location
of the desired and adjacent interfering signals.

In \cite{Jap01} the authors of the present paper proposed a compressed
sensing based \gls{DCR} which uses e.g. prior knowledge in the
signal reconstruction used in compressed sensing to relax the filtering
requirements without significant increase in sampling frequency.
The following sections present further work on this idea.

\section{IRfDUCS project}
\label{sec:irfducs}
Interference Reduction for Direct conversion receiver Using the
Compressed Sensing (IRfDUCS) is a 3 year research project carried out
at Aalborg University, Denmark.\footnote{More information at:
\url{http://www.sparsesampling.com/interference2/pates}.}
As illustrated in Fig.~\ref{fig:system} the project consists of three phases:
1) numerical simulation; 2) hardware implementation; and 3)
integrated circuit design.

\begin{figure*}[t]
            \centering
            \includegraphics{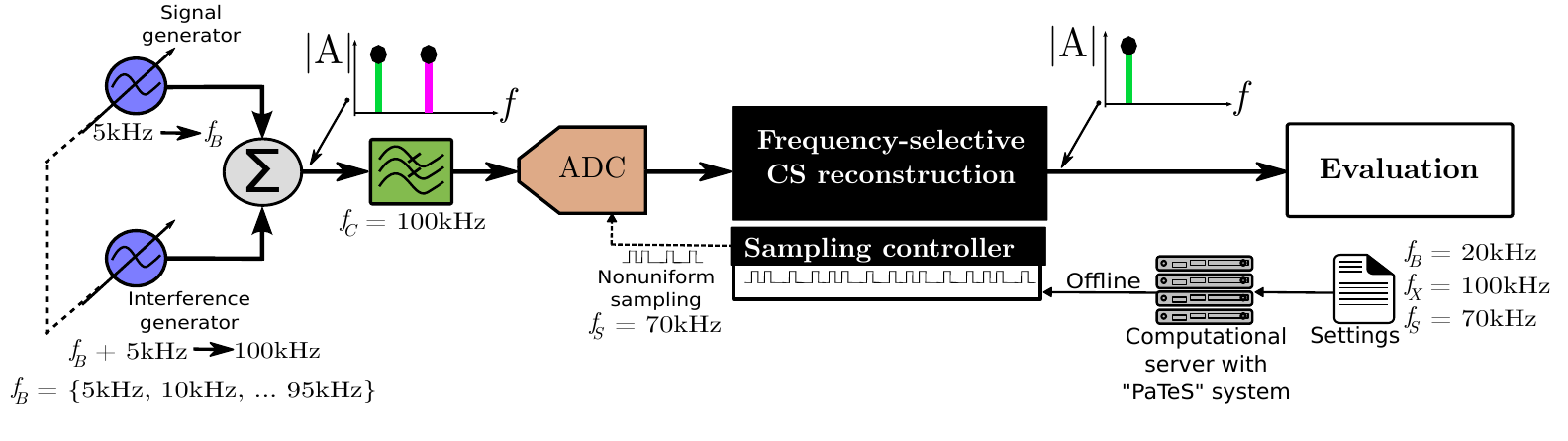}
            \caption{Conducted experiment.}
            \label{fig:system}
\end{figure*}

Currently, the project is in the phase of numerical simulations.
Two software packages are being developed during this phase:
\begin{itemize}
  \item \textbf{RxCS}: Receiver simulation with emphasis on Compressed Sensing.
   This is a software system with a selection of radio signal generators,
   acquisition subsystems, signal reconstruction techniques and
   evaluation modules.
   The software is able to simulate receiver systems
   and compressed sensing systems based on behavioral models.
   It is the main tool used for numerical simulations in the project.
  \item \textbf{PaTeS}: Patterns Testing System.
   This software is dedicated to generate and analyze sampling patterns.
   The most important part of the software performs a quality
   assessment of the compressed sensing reconstruction
   when using different sampling patterns.
\end{itemize}
The second phase of the project targets hardware implementation of the
proposed architecture using \gls{COTS} components. Programmable development
platforms equipped with \glspl{FPGA} and \gls{RISC} devices are to be used in
this phase.

In the third and last phase the developed system is planned to be
designed as an \gls{IC} using a dedicated development tool. The design is to
be performed on schematic and layout level.

\section{System architecture}
\label{sec:arch}

\begin{figure*}[!]
            \centering
            \includegraphics{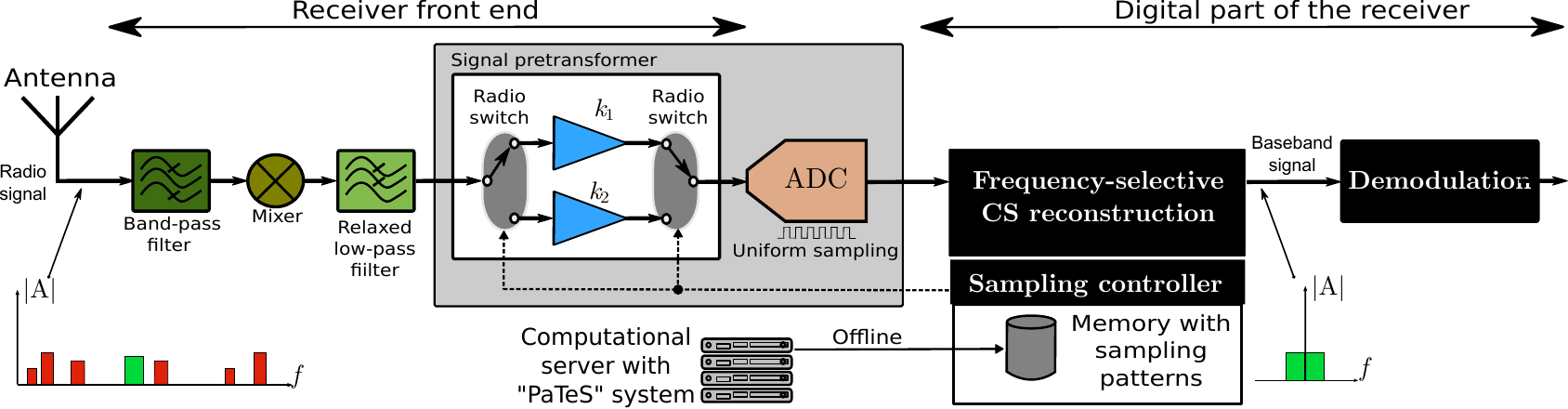}
            \caption{Proposed receiver architecture.}
            \label{fig:arch}
\end{figure*}

The proposed system architecture is presented in Fig.~\ref{fig:arch}. The
received radio signal is filtered by a relatively wide bandpass filter
allowing passage of all relevant channels. Next, this signal is mixed with an
\gls{LO} signal to form a down-converted signal and whatever undesired
interfering signals that are also present after the bandpass filter. Normally
a quadrature down-converter (quadrature-mixer) is used to extract in-phase and
quadrature-phase signal components -- in the presented architecture in
Fig.~\ref{fig:arch} the signal extraction is done in the signal processing
part. As seen from Fig.~\ref{fig:spectrum}, the down-converted signal consists
of a wanted signal (green band) and unwanted interfering signals which must be
removed or at least significantly reduced (red bands). The signal is then
filtered by a low-pass filter which only partially removes the interfering
signals. The loosely-filtered baseband signal is then processed by an
amplifier, and sampled by a uniformly clocked \gls{ADC}. The clock frequency
can be lower than the Nyquist frequency of the baseband signal. The
amplification stage has \gls{AGC} functionality for signal conditioning.

The precise sampling scheme used in the compressed sensing signal
reconstruction depends on several factors such as frequency band
of the desired signal, frequencies of the interferers, sampling frequency etc.
All this is pre-computed by the \texttt{PaTeS} software which makes the
extensive signal condition analysis to allow the real-time processing to
use the best possible sampling scheme.  It is possible to adapt the
scheme to \gls{SDR} \cite{Mit93} compliance by having \texttt{PaTeS} run offline
and downloading new data for the sampling scheme in case for example
the bandwidth of the desired signal changes or similar.  However, normally
it is possible to pre-compute many scenarios but there is the possibility to update
in case there are some special use-cases that are typical or to handle special
situations. This updating is thus optional and provides an added bonus in case
it is used.

The \texttt{PaTeS} software is able to define which strategy to
apply in certain situations -- for example if a sample should be removed if it
allows decreasing the risk of \gls{ADC} saturation or keep the sample if the
signal reconstruction improves significantly by that. This is part of the
compressed sensing technique where loosing a limited number of samples may be
possible without severe effects on the reconstructed signal. The
\texttt{PaTeS} system thus analyzes the conditions for the received signal and
seeks to jointly optimise the reconstruction quality of the desired signal
while relaxing the lowpass filter requirements. The reconstructed signal is
then given to a demodulation module as seen from Fig.~\ref{fig:arch}.

The proposed architecture allows for using relaxed lowpass filter requirements
compared to a traditional \gls{DCR}, without using a fast-clocked \gls{ADC}.
Unlike the previously proposed solution \cite{Jap01}, the \gls{ADC} is clocked
uniformly. The requirements for the dynamic range of the \gls{ADC} are relaxed
-- hence there is some tolerance for saturation of the signal. Instead, the
architecture includes an \gls{AGC} enabled amplifier combined with the
compressed sensing signal reconstruction.

\section{Illustrative example}
\begin{figure}[b]
            \centering
            \includegraphics{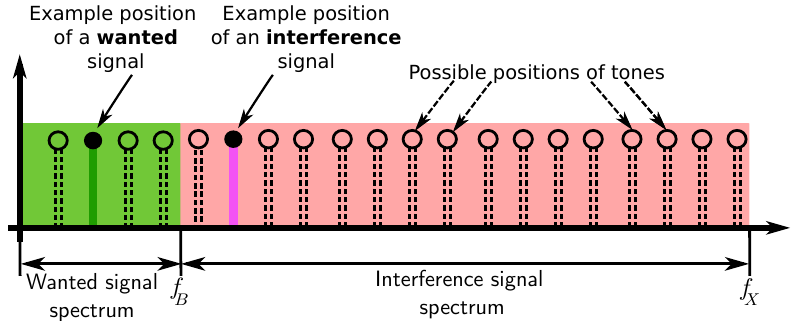}
            \caption{Illustration of the power spectral density after down-conversion.}
            \label{fig:spectrum}
\end{figure}

A numerical experiment was conducted to demonstrate the idea of the project.
In the experiment a sampling pattern was designed for a given desired and
interfering signal constellation. A very simple case of a single tone was used
for both the desired and the interfering signal. The frequencies were
initially chosen as shown in Fig.~\ref{fig:system} with
$f_\mathrm{B}=20\:$kHz. The sampling pattern was designed according to the
setup shown in Fig.~\ref{fig:system}. Both frequencies were varied in
increments of $5\:$kHz to see the effect of frequency changes. During the
simulation the frequency of the desired frequency was swept from $5\:$kHz to
$95\:$kHz -- thus significantly above the maximum frequency for which the
sampling pattern was generated. The purpose was to see the effect of violating
the design constraint. The power spectral density of the signals is
illustrated in Fig.~\ref{fig:spectrum}.

\begin{figure}[b]
            \centering
            \includegraphics{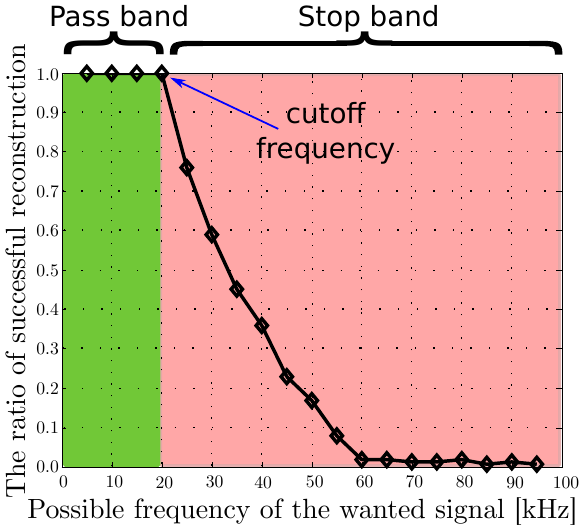}
            \caption{Simulated probability of successful desired signal reconstruction versus
            frequency of the desired signal.}
            \label{fig:res}
\end{figure}

Fig. \ref{fig:res} shows the probability of successful signal reconstruction
of the wanted signal versus the frequency of the desired signal. As seen the
signal is perfectly reconstructed for all frequencies up to the maximum design
frequency of $20\:$kHz. When the frequency of the desired signal increases
above the design frequency of $20\:$kHz the probability of correct signal
reconstruction reduces as expected and above $55\:$kHz it is impossible to
obtain correct reconstruction. This shows that the principle works and that
the interference signal can be handled within the design constraint of the
sampling pattern.

\section{Conclusions}
A modified \gls{DCR} architecture allowing relaxed filtering requirements
compared to a standard \gls{DCR} has been presented. The proposed architecture
was based on adapting the compressed sensing technique to utilize two aspects:
1) prior knowledge of frequency placement of the desired and interfering
signals; and 2) selecting the sampling patterns according to the placement of
the interfering signals. Combined the concepts allowed a compressed sensing
formulation which was tailored to relax the analog filtering requirements
while meeting the quality metrics for the reconstructed in-phase and
quadrature-phase signals. A proof of concept has been demonstrated in a simple
configuration and research is ongoing to further advance the technique.

\section{Acknowledgment}
The work is supported by The Danish Council for Independent Research under
grant number 0602--02565B. Part of the work was supported by The Danish
National Advanced Technology Foundation under grant number 035-2009-2.

\end{document}